\documentclass{aa}  
\usepackage[fleqn]{amsmath}
\usepackage{physics} 
\usepackage{textcomp}
\usepackage{amsfonts} 
\usepackage{graphicx}
\usepackage{dblfloatfix}
\usepackage[a]{esvect}
\usepackage{bm}
\usepackage{upgreek}
\usepackage{xcolor}
\usepackage{txfonts}
\usepackage[normalem]{ulem}
\usepackage{natbib}
\usepackage{makecell} 
\usepackage{comment}
\usepackage{float}
\usepackage{booktabs}
\usepackage{tabularx}
\usepackage{orcidlink}
\usepackage{siunitx}
\usepackage{hyperref}

\begin{document} 
   \title{Accurate spectroscopic redshift estimation using non-negative matrix factorization: application to MUSE spectra} 
   \author{Masten Bourahma\,\orcidlink{0009-0008-8284-1490}\inst{1}
           \and
           Nicolas F. Bouché\,\orcidlink{0000-0003-0068-9920}\inst{1}
           \and
           Roland Bacon\,\orcidlink{0000-0001-6148-0915}\inst{1}
           \and
           Johan Richard\,\orcidlink{0000-0001-5492-1049}\inst{1}
           \and
           Tanya Urrutia\inst{2}
           \and
           Afonso Vale\,\orcidlink{0009-0008-7166-0512}\inst{3,4}
           \and
            Martin Wendt\,\orcidlink{0000-0001-5020-9994}\inst{5} 
           \and
           T. T. Thai\,\orcidlink{0000-0002-8408-4816}\inst{6}}
    \institute{
      Université Claude Bernard Lyon 1, Ens de Lyon, CNRS, Centre de Recherche Astrophysique de Lyon (CRAL) UMR5574, F-69230 Saint-Genis-Laval, France \\ \email{masten.bourahma@univ-lyon1.fr}
      \and
      Leibniz-Institut für Astrophysik Potsdam (AIP), An der Sternwarte 16, 14482 Potsdam, Germany
      \and
      Instituto de Astrofísica e Ciências do Espaço, Universidade do Porto, CAUP, Rua das Estrelas, PT4150-762 Porto, Portugal
      \and
      Departamento de Física e Astronomia, Faculdade de Ciências, Universidade do Porto, Rua do Campo Alegre 687, PT4169-007 Porto, Portugal
      \and
      Institut f\"{u}r Physik und Astronomie, Universit\"{a}t Potsdam, Karl-Liebknecht-Str. 24/25, 14476 Golm, Germany 
      \and
      National Astronomical Observatory of Japan, 2-21-1 Osawa, Mitaka, Tokyo 181-8588 Japan
    }
   \date{Received ---; accepted ---}
 
  \abstract{Accurate and automated galaxy redshift determination is essential for maximizing the scientific return of spectroscopic surveys. In this paper, we propose a data-driven method to address this challenge. The method first learns a rest-frame representation of galaxy spectra using Non-negative Matrix Factorization (NMF). The method then reconstructs new spectra using this representation at different trial redshifts, and identifies the correct redshift by selecting the one that minimizes the reconstruction error.
  
  We apply our method to galaxy spectra from the Multi Unit Spectroscopic Explorer (MUSE), covering redshifts from 0 to 6.7. Our method achieves an overall success rate of 93.7\%.
  We further demonstrate two applications: (i) the separation between true and false sources, and (ii) the detection of blended sources from one-dimensional spectra. Our results demonstrate that NMF-based representations provide a powerful and physically motivated framework for redshift estimation in current and future large spectroscopic surveys.}

\keywords{spectra, spectroscopic redshift, galaxies, NMF, Machine Learning}
\titlerunning{NMF on MUSE spectra}
\authorrunning{Bourahma, Bouché et al.}
\maketitle
\section{Introduction}
All-sky Multi-Object Spectroscopic (MOS) surveys such as the Sloan Digital Sky Survey \citep[SDSS][]{York_2000} and its upgrades (BOSS, eBOSS) have ushered astronomy into the big-data era, producing millions of spectra for objects down to 18–19 mag. MOS surveys on $4$ meter class telescopes, including the Dark Energy Spectroscopic Instrument \citep[DESI][]{DESI_Collaboration_2022}, the $4$-Meter Multi-Object Spectroscopic Telescope \citep[4MOST][]{2019Msngr.175....3D}, and the WHT Enhanced Area Velocity Explorer \citep[WEAVE][]{WEAVE}, now deliver tens of millions of spectra reaching $\sim22$ magnitude. On a larger telescope, the Multi-Object Optical and Near-infrared Spectrograph \citep[MOONS][]{moons} at the VLT will extend this capability into the near-infrared. To maximize the scientific outcome of these large data surveys, strategies to perform fast, efficient, and accurate automated object classification and redshift identification have been developed.

State-of-the-art tools for automated galaxy redshift estimation rely on spectral templates fitting, where a basis of galaxy spectra is first constructed and subsequently used to match observations across trial redshifts. These tools adopt different techniques and strategies to build a set of representative templates of the galaxy population. For example, \citep{2012AJ....144..144B} uses a Principal Component Analysis (PCA) procedure that accounts for measurement errors and missing data to learn a rest-frame representation of SDSS galaxies. Similarly, \texttt{AUTOZ} \citep{AUTOZ}, $\texttt{MARZ}$ \citep{HINTON201661, Inami_2017}, and \texttt{xPCA} (Krogager, in prep) use cross-correlation with a set of data-driven templates obtained with PCA, where stellar continuum is subtracted. \texttt{Redrock} \citep{RedRock} developed for the DESI pipeline uses a set of synthetic archetype galaxy templates that were obtained from stellar population synthesis and models of emission line fluxes. Deep learning approaches have also been explored to derive rest-frame representations. For instance, \citet[4MOST]{GASNET_III} trained an encoder network "\texttt{GaSNet-III}" to learn a rest-frame representation of SDSS galaxy spectra and used it for redshift prediction, achieving competitive accuracy. Other deep learning models that do not rely on a rest-frame representation have also been explored. \citet{GASNET_III} trained a U-net network to transform spectra from the observed to the rest frame directly. Another example of such approaches is \texttt{M-TOPnet} \citep[MOONS]{M-TOPnet}. This network was trained through a multi-task learning framework to simultaneously predict the redshift probability density function, stellar mass, star formation rate, and emission-line locations from the continuum-subtracted spectrum and a condensed continuum vector. All of these tools demonstrated speed, high accuracy, and robustness within their respective MOS survey contexts.

While MOS surveys cover relatively bright pre-selected objects from imaging surveys, Integral Field Spectroscopy (IFS) instruments, such as the Multi Unit Spectroscopic Explorer \citep[MUSE][]{BaconR2010}, provide spectra of all objects in the field, i.e., without pre-selections, down to magnitudes $28$ and beyond \citep{BaconR2017,BaconR2023}. At the same time, MUSE poses specific challenges because of its wavelength coverage resulting in objects detected over a wide redshift range ($z = 0-6.7$), which can lead to emission line confusion, especially between [\ion{O}{ii}] $\lambda\lambda3727,3729$\,\AA\ at $z<1.5$ and Ly$\alpha$ $\lambda1216$\,\AA\ at $z>2.8$.
Thanks to its exquisite sensitivity, MUSE can efficiently find emission line objects without any continuum.
As a result, the task of source detection and redshift identification in deep fields needs to be performed twice \citep{BaconR2023}, once on continuum-detected objects and once on emission line-detected objects, and currently, no existing tool can perform well on both types without visual inspections and validation. 
In addition, in the "redshift desert" at $1.5 < z < 2.8$, galaxies lack strong spectral emission features, making redshift determination heavily dependent on continuum shape. These challenges call for new approaches capable of handling both continuum and emission-line–dominated spectra across the full MUSE redshift range.

We present a method for automated galaxy redshift prediction, enabled by the availability of approximately 10,000 MUSE spectra with redshift labels. Our approach is based on Non-negative Matrix Factorization (NMF), which learns a low-rank, additive, and non-negative representation of galaxy spectra. By enforcing non-negativity, NMF provides a parts-based and more interpretable representation compared to other techniques like PCA. Moreover, NMF has been successfully applied in astrophysics to construct empirical spectral templates \citep{BlantonM2007a} and to identify fundamental mid-infrared components of galaxy spectra from Spitzer/IRS \citep{Hurley_2013}, demonstrating its effectiveness. The method finds the best redshift solution by non-negatively projecting a spectrum onto this representation for a range of trial redshifts, and then selecting the projection with the lowest reconstruction error. We apply our method to an independent test set and show its robustness.

This paper is organized as follows. In~Sect.~\ref{sec:data}, we describe the data selection and preprocessing procedures. Sect.~\ref{sec:method} presents the application of NMF to MUSE galaxy spectra and the general methodology underlying our redshift prediction framework. In~Sect.~\ref{sec:application}, we report the test results and demonstrate two applications of the method: (i) false and true sources separation, and (ii) spectral deblending. Finally, we discuss our results in~Sect.~\ref{sec:discussion} and conclude in~Sect.~\ref{sec:conclusions}.
\section{Data} 
\label{sec:data}
In this work, we used a collection of galaxy spectra taken from five MUSE Guaranteed Time Observations (GTO) surveys:
(i)  the MUSE Hubble Ultra-Deep Field (HUDF) surveys \citep[][PI: R. Bacon]{BaconR2023}, 
(ii) the MusE GAs FLOw and Wind [MEGAFLOW] survey \citep[][PI: N. Bouché]{Bouch__2025}, 
(iii) the MUSCATEL survey (Urrutia, in prep; PI: L. Wisotzki),
(iv) the MUSE-WIDE DR1 survey \citep[][PI: L. Wisotzki]{Urrutia_2019}, and the MUSE gAlaxy Groups in COSMOS (MAGIC) survey \citep[][PI: T. Contini]{Epinat_2024}. 

The combination of these surveys provides a representative and diverse sample of galaxy spectra for learning and analysis. It includes galaxies of different types and spectral characteristics—from continuum-dominated systems to strong emission- and absorption-line galaxies—spanning wide redshift and magnitude ranges. The mixture of deep and shallow fields observed under different conditions provides spectra with varying signal-to-noise ratios (SNR), enabling us to evaluate the robustness of our methods under realistic observational conditions.  Table \ref{tab:surveys_table} summarizes some of the key characteristics of each survey.

From these surveys, we selected spectra of galaxies based on two criteria. First, we required a secure redshift by selecting galaxies with a redshift confidence score (ZCONF) of 1 or higher~\footnote{
As  discussed in \citet{BaconR2023}, 
redshifts were obtained using a semi-automatic procedure where the five best redshift solutions produced by a MUSE-adapted version of the \texttt{MARZ} algorithm were visually inspected by a group of experts, who evaluated the reliability of the redshift on a scale from 0 to 3: a ZCONF of 3 indicates a secure redshift from multiple spectral lines; 2 represents a highly probable redshift from a single high SNR line or from multiple low SNR lines; 1 denotes a possible redshift from a low SNR line; and 0 is attributed when the redshift is undetermined.}. Second, we visually inspected the spectra to exclude most of the bright blended spectra "blends" (overlap between multiple sources). 

\begin{figure}
    \centering
    \includegraphics[width=1\linewidth]{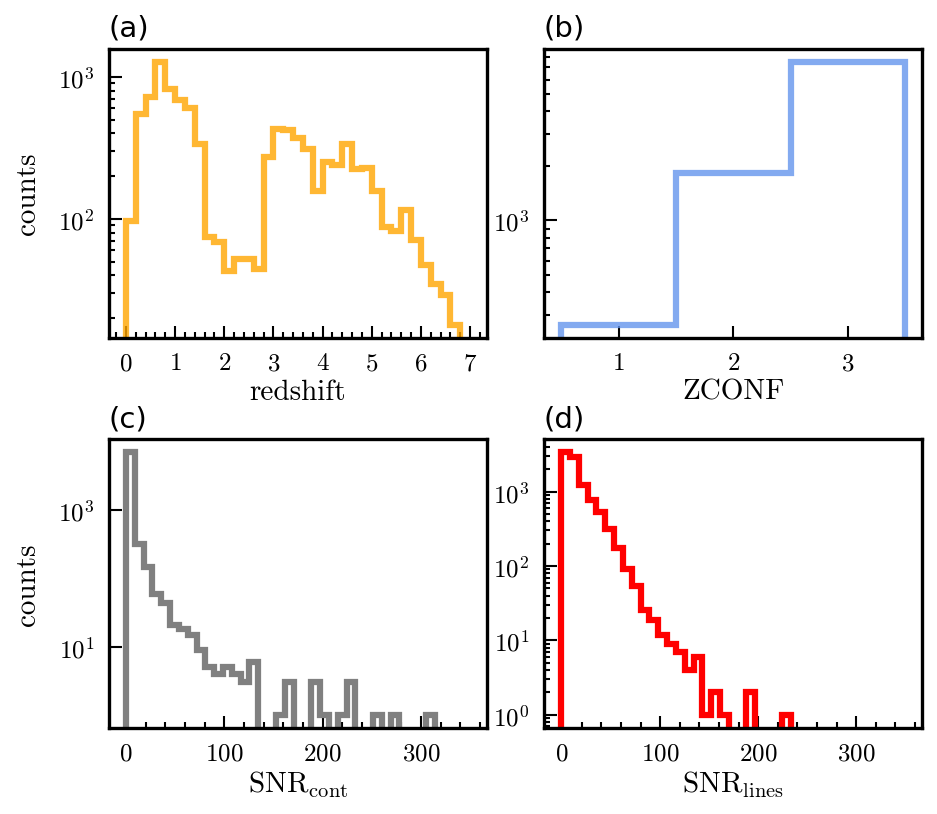}
    \caption{ Main statistical properties of the selected MUSE galaxy sample. Distributions of (a) redshift, (b) redshift confidence score (ZCONF), (c) stellar continuum SNR ($\rm SNR_{cont}$), and (d) emission/absorption lines collective SNR ($\rm SNR_{lines}$).}
    \label{fig:statistics}
\end{figure}

Our selection yielded a sample of \num{9252} spectra. Fig.~\ref{fig:statistics} summarizes the main properties of the sample, showing the distributions of redshift, ZCONF, continuum SNR ($\rm SNR_{cont}$), and lines' SNR ($\rm SNR_{lines}$). We computed $\rm SNR_{cont}$ as the median SNR of the stellar continuum in the observed-frame \num{7650}-\num{7850}\,\AA\ window, chosen for its location within the region of highest MUSE throughput. In a given spectrum, we computed $\rm SNR_{lines}$ as the quadrature sum of the signal-to-noise ratios of all detected emission and absorption lines with detection significances exceeding \num{3}$\,\sigma$. The stellar continuum fit and line measurements were performed with the \texttt{pyPlatefit} package \citep{BaconR2023}, specifically developed for MUSE spectra. The preprocessing steps specific to our method are described in the following section.

\begin{table}
\centering
\caption{Summary of the MUSE surveys used in this work.}
\small
\begin{tabularx}{\columnwidth}{@{}>{\raggedright\arraybackslash}p{1.9cm} |p{2.75cm}|c|l @{}}

\textbf{Survey} & \textbf{Area \& depth} & \textbf{\# Spectra} & \textbf{Ref.} \\ [2pt]
\hline
\textbf{MUSE HUDF} & & & \\ 
\quad UDF10 & $1'\times1'$ field, $10$\,h & 351 & (1) \\ 
\quad MOSAIC & $3'\times3'$ field, $31$\,h & 1505 & (1) \\
\quad MXDF & $1'$-diameter circular field, $141$\,h deep & 691 & (1) \\[3pt]

\textbf{MEGAFLOW} & 22 quasar fields: 20 shallow ($\sim$3\,h) and 2 deep ($\sim$11\,h) & 2427 & (2) \\

\textbf{MAGIC} & 17 fields in COSMOS, depths between 1–10\,h & 1423 & (3) \\

\textbf{MUSE-WIDE DR1} & 100 pointings in GOODS-S and CANDELS, each 1\,h deep & 1602 & (4) \\

\textbf{MUSCATEL} & Deep (25\,h), medium (5\,h) and shallow (100 \,min) observations in the Frontier Field Parallels. & 4545 & (5) \\ 
\end{tabularx}
\tablebib{(1)~\citet{BaconR2023}; (2)~\citet{Bouch__2025}; (3)~\citet{Epinat_2024}; (4)~\citet{Urrutia_2019}; (5)~Urrutia, in prep.}
\label{tab:surveys_table}
\end{table}
\section{Methods}
\label{sec:method}

\subsection{Non-negative matrix factorization}
Non-negative matrix factorization is a Machine Learning technique for dimensionality reduction \citep{Lee1999}. Given a data matrix $\bm{X}$, NMF learns a low-dimensional representation by approximating $\bm{X}$ as the product of two non-negative, low-rank matrices, $\bm{W}$ and $\bm{H}$
\begin{equation}
\label{eq:nmf_1}
\begin{aligned}
 &  \quad \quad \quad \bm{X} \simeq \bm{W} \bm{H}, \\
 &   \quad \quad \quad \bm{X} \in \mathbb{R}_+^{n \times l}, \bm{W} \in \mathbb{R}_+^{n \times k}, \bm{H} \in \mathbb{R}_+^{k \times l}, \text{ and } k << n, l.
\end{aligned}
\end{equation}

\noindent In the specific case of galaxy spectra arranged row-wise in $\bm{X}$ (see, Fig.~\ref{fig:spectra_matrix}), NMF reconstructs each spectrum in $\bm{X}$ as a positive linear combination of $k$ basis vectors in $\bm{H}$ with coefficients placed in the corresponding row of $\bm{W}$. The number of basis vectors is directly controlled by the rank $k$; its value is free and must be fine-tuned (refer to \ref{subsec:kft}). \\

The prevalent algorithm to solve Eq.~\ref{eq:nmf_1} is described in the seminal work of \citeauthor{Lee1999}. For our case, we adopt the more sophisticated "nearly-NMF" algorithm developed by \citet{GreenBaily24} specifically for astronomical spectra. nearly-NMF extends \citet{Lee1999} and \citet{ZhuG_16a} algorithms to account for heteroscedastic uncertainties and missing values. It also considers negative values in $\bm{X}$, known to cause offsets in basis vectors if zero-clipped \citep{GreenBaily24}. Briefly, they considered the following optimization problem,
\begin{equation}
\label{eq:objective}
\begin{aligned}
& \quad \quad \quad \min_{\bm{W}\in \mathbb{R}^{n \times k}, \bm{H}\in \mathbb{R}^{k \times l}} &&||(\bm{X+Y}) - (\bm{WH + Y})||_{\bm{F}}^2\\
& \quad \quad \quad \text{\quad \quad s.t.} && \bm{W} \geq 0 \;\text{and} \; \bm{H} \geq 0, 
\end{aligned}
\end{equation}
where matrix $\bm{Y}$ \footnote{matrix $\bm{Y}$ changes during iterations and can be different for $\bm{W}$ and $\bm{H}$ updates} holds the minimum shift to make each entry in $\bm{X}$ non-negative. $\bm{F}=1 /\bm{\sigma}^2$ is the inverse-variance matrix associated with $\bm{X}$, a zero weight is assigned to missing entries. To minimize this objective, they derived multiplicative iterative update rules for $\bm{W}$ and $\bm{H}$ matrices that guarantee convergence to a local minimum,

\begin{equation}
   \begin{array}{l}
      \bm{H} = \bm{H} \odot \dfrac{\left[ \bm{W}^{T} (\bm{V} \odot \bm{X}) \right]^{+}}{ \bm{W}^{T}(\bm{V} \odot (\bm{WH})) + \left[\bm{W}^{T}(\bm{V} \odot \bm{X}) \right]^{-}}\\
      \bm{W} = \bm{W} \odot \dfrac{\left[ (\bm{V} \odot \bm{X})\bm{H}^{T} \right]^{+}}{ (\bm{V} \odot (\bm{WH}))\bm{H}^{T} + \left[(\bm{V} \odot \bm{X})\bm{H}^{T} \right]^{-}} \; ,\\
   \end{array}
   \label{eq:nmf_updates}
\end{equation}
where $\odot$ is the element-wise product. Operation $[\;]^{+}$ applied on a matrix results in a matrix of the same shape, in which all negative values are zeroed. $[\;]^{-}$ on the other hand, zeroes the positive values and takes the absolute value of negative values.

Next, we describe how we construct the matrix $\bm{X}$ and the related processing steps. We then explain (in~Sect.~\ref{subsec:redshifts}) how we can predict redshift using NMF learned basis vectors, and present several metrics used to validate the predictions (in ~Sect.~\ref{subsec:metrics}). Finally, we present the rank selection and validation methods in~Sect.~\ref{subsec:kft}.

\subsection{Data transformation}
\label{subsection:data_transformation}
\begin{figure}
    \centering
    \includegraphics[width=1\linewidth]{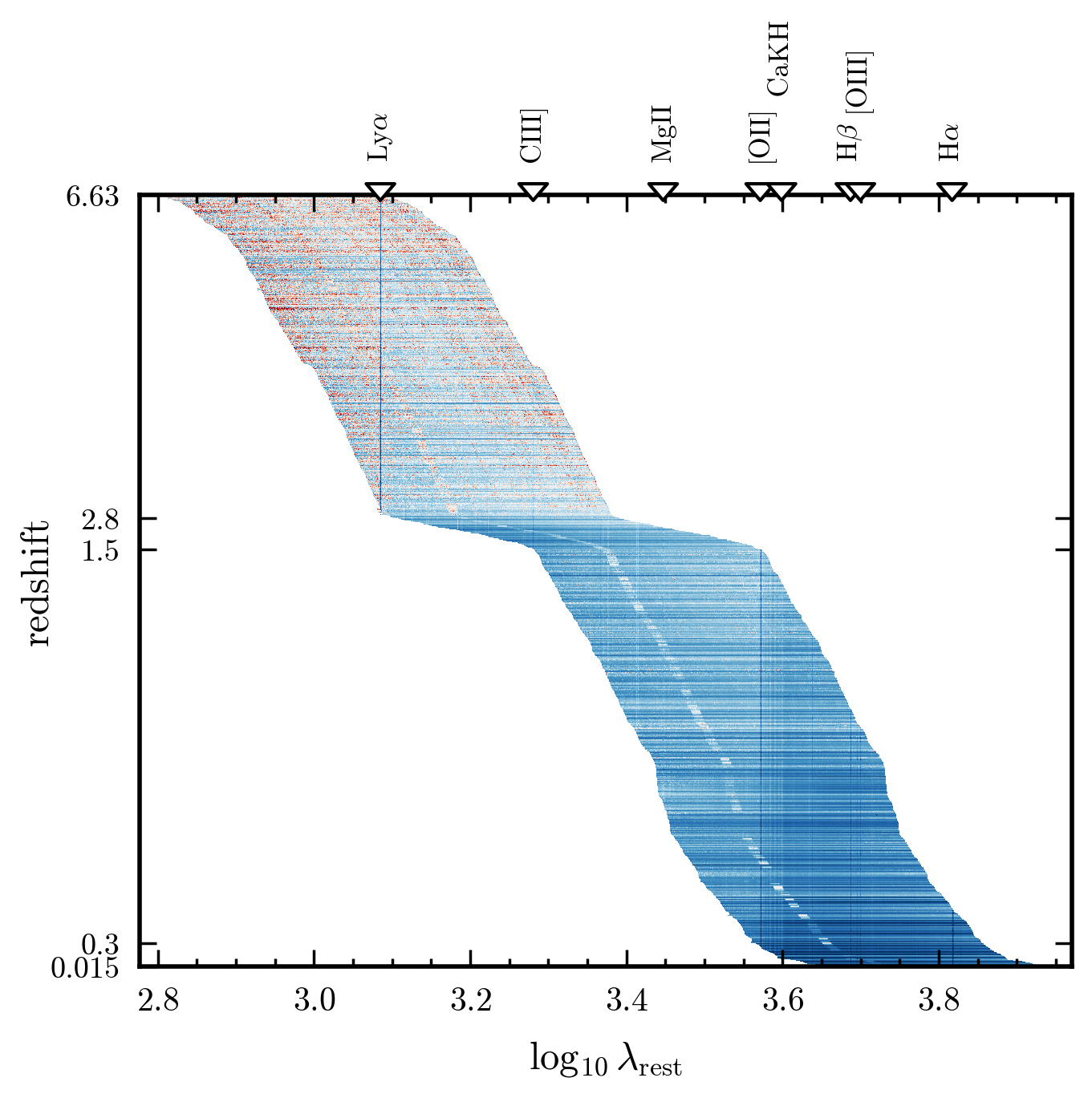}
    \caption{MUSE galaxy spectra matrix in the rest frame. This matrix shows selected MUSE galaxy spectra, prepared for NMF decomposition. Spectra are sorted by increasing redshift (from bottom to top) and transformed to their rest frame. Key redshifts are shown on the y-axis, and important spectral lines are indicated on top of the figure. The color of each pixel encodes flux density, scaled using a 95\% z-scale to enhance the visibility of emission lines. Bluer colors represent higher flux densities; white pixels denote missing or unobserved data.}
    \label{fig:spectra_matrix}
\end{figure}

The rest frame is more effective for learning a representation of galaxy spectra with NMF, as spectral features align consistently at fixed wavelengths, enabling their learning. Consequently, we transform our MUSE galaxy spectra sample to the rest frame following these steps,

\begin{enumerate}
    \item We define a common rest-wavelength grid, $\Gamma$, in logarithmic space. This grid spans the range $2.77 \leq \log_{10}{\lambda} \leq 3.97$, with a uniform logarithmic spacing $\Delta \log_{10}{\lambda}=2.215525 \times 10^{-5}$. The lower end of the grid corresponds to the bluest MUSE wavelength ($\lambda_{\rm obs, min} = 4600\,\mathring{\rm A}$), and transformed into the rest frame assuming a maximum redshift of $z_{\rm max} = 6.7$. The upper end of the grid is given by the reddest MUSE wavelength ($\rm \lambda_{obs,max} = 9350\,\mathring{\rm A}$) at redshift zero.
    We derived the grid's spacing, $\Delta \log_{10}{\lambda}$, from an equivalent reference linear grid. This linear grid spans the same spectral range and has a step size, $\rm \Delta \lambda_{ref}$, determined by the maximally squeezed MUSE spectral sampling ($\delta \lambda$) in the rest frame, 
    \begin{equation*}
        \Delta \lambda_{\rm ref} = \dfrac{\delta\lambda}{1 + z_{\rm max}} = \dfrac{1.25}{1 + 6.7} = 0.16233\,\mathring{\rm A}.
    \end{equation*}
    When this linear grid is transformed directly into logarithmic space, it produces non-uniform logarithmic spacings. $\Delta \log_{10}{\lambda}$ is then taken as the mean of these resulting logarithmic spacings to define a constant. These choices resulted in a 53918-dimensional grid. \\
    
    \item We express flux densities $f_{\lambda}$ in terms of the logarithmic rest wavelengths
    \begin{equation}
    \begin{aligned}
        & f_{\lambda}(\log_{10} \lambda_{\rm rest}) = \ln{10}\,f_{\lambda}(\lambda_{\rm obs}) \lambda_{\rm obs}, \\
        & \log_{10} \lambda_{\rm rest} = \log_{10} \lambda_{\rm obs} - \log_{10} (1 + z),
    \end{aligned} 
    \label{eq:wavelength_transform}
    \end{equation}
    where $\lambda_{\rm rest}$, $\lambda_{\rm obs}$ are respectively the rest and observed wavelengths, and $z$ is the redshift. \\
    
    \item For each spectrum, we linearly interpolate the pairs  $\left( f_{\lambda}(\log_{10} \lambda_{\rm rest}), \; \log_{10} \lambda_{\rm rest} \right)$ to the rest-frame wavelength grid $\Gamma$, and extrapolate with zeros in unobserved regions of  $\Gamma$. Variances undergo the same transformations, except for the extrapolation part, where we assign a large number instead of zero. 
\end{enumerate}
Figure~\ref{fig:spectra_matrix} shows the result of the rest-frame transformation of our sample of galaxy spectra.

\subsection{Estimating redshift using NMF basis vectors}
\label{subsec:redshifts}
Our method posits that the correct rest-frame of NMF basis vectors---aligned with the true redshift $z_t$---achieves the best representation for a given observed galaxy spectrum. Conversely, an incorrect rest-frame will result in a poorer representation. For instance, consider a spectrum exhibiting nebular emission lines such as \mbox{[\ion{O}\,\textsc{ii}]}, H$\beta$, strong and weak \mbox{[\ion{O}\,\textsc{iii}]}. Only the correct rest-frame allows for a simultaneous and accurate reconstruction of all these features. In contrast, an incorrect rest-frame may still reconstruct individual emission lines but will fail to model all of them collectively, as they are intrinsically absent. \\

\begin{figure*}[t]
    \centering
    \includegraphics[width=1\linewidth]{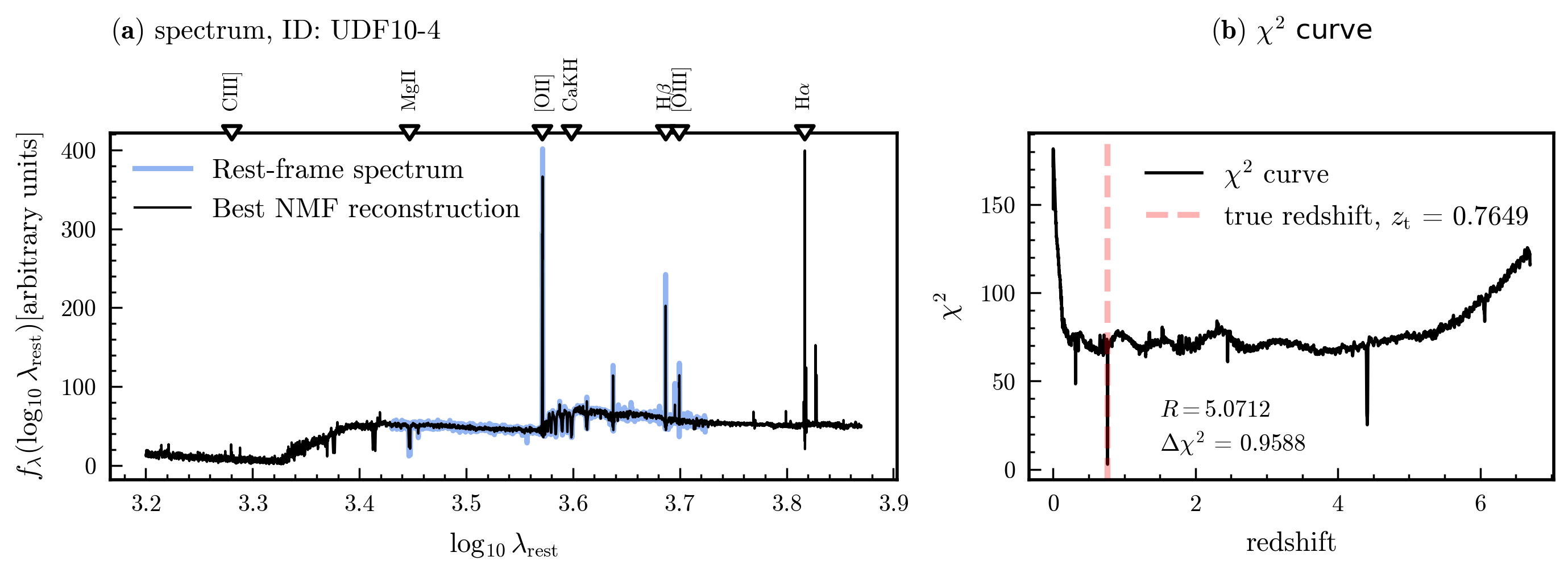}
    \caption{Illustration of redshift prediction with NMF basis vectors. (a) shows the rest-frame spectrum of the UDF10-4 source at redshift 0.7649 in blue. This galaxy exhibits stellar continuum, strong \mbox{[\ion{O}\,\textsc{ii}]}, H$\beta$, and \mbox{[\ion{O}\,\textsc{iii}]} spectral emission lines (their locations are indicated on top of the figure); the best NMF reconstruction is plotted in black. (b) reports the corresponding $\chi^2$ curve, the minimum happens at the true redshift (vertical red dashed line), hence, successfully predicting the correct redshift. The second minimum in the $\chi^2$ curve at $z\sim$\,4.4 corresponds to a solution in which \mbox{[\ion{O}\,\textsc{ii}]} gets mistaken to be Lya. The values of the $\Delta \chi^2$ and $R$ metrics are also reported in (b); their values indicate a significant minimum and a very robust redshift prediction.}
    \label{fig:method}
\end{figure*}

Following this principle, to determine the redshift of an observed spectrum, $\vec{f_{\rm \lambda, \,obs}}$, we test all redshifts in the $\rm 0 \leq z \leq 6.7$ range with a step of \num{0.0005}. For each test redshift, $z_{\rm test}$, we proceed as follows:

\begin{enumerate}[(a)]
    \item We de-redshift the spectrum to its rest-frame assuming $z_{\rm test}$, following the same steps in \ref{subsection:data_transformation}, we note the de-redshifted spectrum $\vec{f_{\rm \lambda, \,test}}$. \\
    \item We reconstruct $\vec{f_{\rm \lambda, \,test}}$ using NMF basis vectors,
    \begin{equation}
        \vec{f_{\rm \lambda, \,test}} \simeq \vec{\omega}^T \bm{H},
        \label{eq:nmf_reconstruction}
    \end{equation}
    vector $\vec{\omega}$ contains decomposition coefficients. Eq.~\ref{eq:nmf_reconstruction} is solved using a Non-Negative Least Squares \citep{fnnls}, ensuring that the coefficients are non-negative as in NMF.\\
    \item We assess the reconstruction quality at $z_{\rm test}$ by computing the corresponding $\chi^2$ goodness-of-fit statistic,
    \begin{equation}
        \chi^2(z_{\rm test}) = \displaystyle \sum_{i=1}^{L} \left( \dfrac{\vec{f_{\rm \lambda, \,test}} -  \vec{\omega}^T \bm{H}}{ \vec{\sigma_{\rm \lambda, \,test}}} \right)_i ^2,
        \label{eq:nmf_chi2test}
    \end{equation}
    $\vec{\sigma_{\rm \lambda, \,test}}$, is the corresponding standard deviations vector of $\vec{f_{\rm \lambda, \,test}}$. $L$ is the spectral dimension. 
\end{enumerate}
Testing all redshifts results in a $\chi^2(z)$ curve. The minimum of this curve $\chi^2_{\rm min}$ then gives the predicted redshift $z_p$. Moreover, we characterize this minimum with a $\Delta \chi^2$ and a robustness score (see next subsection for definitions). We illustrate our method for a galaxy spectrum taken from the MXDF survey in Fig.~\ref{fig:method} (a), and we show the corresponding $\chi^2$ curve in (b).

\subsection{Evaluation Metrics}
\label{subsec:metrics}

To evaluate a redshift prediction, we used the following quantities from the $\chi^2(z)$ curve:
\begin{enumerate}[(a)]
    \item Redshift significance score~\footnote{\label{fn:definitions}Note that our definitions for the redshift significance and robustness scores differ from other definitions that can be found in SDSS or DESI literature.}
    \begin{equation}
        \Delta \chi^2 = 1-\dfrac{\chi^2_{\rm min}}{\rm Q_1(\chi^2)},
        \label{eq:nmf_dchi2}
    \end{equation}
    where $\chi^2_{\rm min}$ denotes the minimum of the $\chi^2$ curve and $\rm Q_1(\chi^2 )$ its first quartile. The first quartile provides a robust estimate of the typical  $\chi^2$ away from the minimum (baseline of the $\chi^2$ curve).

    Thus, the quantity $\Delta \chi^2$ measures how much the minimum deviates from the baseline of the $\chi^2$ curve, with higher values reflecting more significant minima.\\
    
    \item Redshift robustness score~\footref{fn:definitions}
    \begin{equation}
            R = \rm \dfrac{\hbox{min}^{(2)}{(\chi^2)} - \hbox{min}^{(1)}{(\chi^2)}}{\sigma_{Q1}(\chi^2)}.
            \label{eq:nmf_robustness}
    \end{equation}
    $\rm min^{(1)}{(\chi^2)}$ and $\rm min^{(2)}{(\chi^2)}$ denote the first and second minima of the $\chi^2$ curve, respectively. The quantity $\rm \sigma_{Q1}(\chi^2)$ is the standard deviation of all $\chi^2$ values less than or equal to the first quartile. It provides an estimate of the intrinsic dispersion of points lying on or below the $\chi^2$ curve baseline. Normalizing the separation between the first and second minima by $\rm \sigma_{Q1}(\chi^2)$ measures its significance relative to the typical fluctuations of the $\chi^2$ curve, while being insensitive to high-$\chi^2$ fluctuations. As the two minima approach each other, the score tends toward zero, indicating confusion between two redshift solutions.
\end{enumerate}

To test our method on a set of spectra with known redshifts, we use the following metrics from the predicted and true redshifts:
\begin{enumerate}[(a)]
    \item Relative error
    \begin{equation}
      \delta z = \dfrac{|z_{\rm p} - z_{\rm t}|}{ 1 + z_{\rm t} },    
    \end{equation}
    where $z_{\rm p}$ is the predicted redshift and $z_{\rm t}$ the true redshift.\\
    
    \item The Good Fraction (GF)
    
    This is the ratio of the number of good redshift predictions $N_{\rm good}$ to the total number of predictions $N_{\rm total}$. A redshift prediction is good when its error $\Delta z = |z_{\rm p} - z_{\rm t}| < t_{\rm MUSE} \left(1 + z_{\rm t}\right)$, where $t_{\rm MUSE}$ is the adopted error-tolerance threshold. In this work, we fix $t_{\rm MUSE}$ at 0.005, which corresponds to a value ten times the step size in the trial redshifts.
    \begin{equation}
        {\rm GF} = \dfrac{N_{\rm good}}{N_{\rm total}}. 
    \end{equation}

    \item Mean Absolute error with outlier rejection
    \begin{equation}
    \mathrm{MAE} = \dfrac{1}{N}\sum_{i=1}^{N} \left(\dfrac{|z_{\rm p} - z_{\rm t}|}{1 + z_{\rm t}} \right).
    \end{equation}
    Redshift predictions with a relative error five times larger than the Median Absolute deviation (MAD) are rejected. The MAD is defined as,
    \begin{equation*}
        \mathrm{MAD} = \mathrm{median}\left(\left|\vec{\delta z} - \mathrm{median}(\vec{\delta} z)\right|\right), 
    \end{equation*}
    where $\vec{\delta} z$ is the vector of redshift relative errors.
\end{enumerate}
\subsection{Rank selection and validation of the method}
\label{subsec:kft}
The factorization rank $k$ sets the number of basis vectors in the NMF decomposition and therefore controls the level of detail the model can represent. If $k$ is too large, the model may adjust to noise; whereas if it is too small, it may fail to capture important spectral features. In both cases, the resulting basis vectors generalize poorly, making careful rank selection essential.

In this work, we propose a rank selection strategy based on the performance in the redshift prediction task. We employ a K-fold cross-validation scheme \citep{kohavi}, in which the data is partitioned into K folds, with K-1 folds used to learn basis vectors under various rank hypotheses, and performance is evaluated on the withheld fold. This process is repeated until each of the K folds has served as the testing set once. The best rank is then taken to be the one that provides the best performance across all testing folds.

\begin{figure}[h]
    \centering
    \includegraphics[width=1\linewidth]{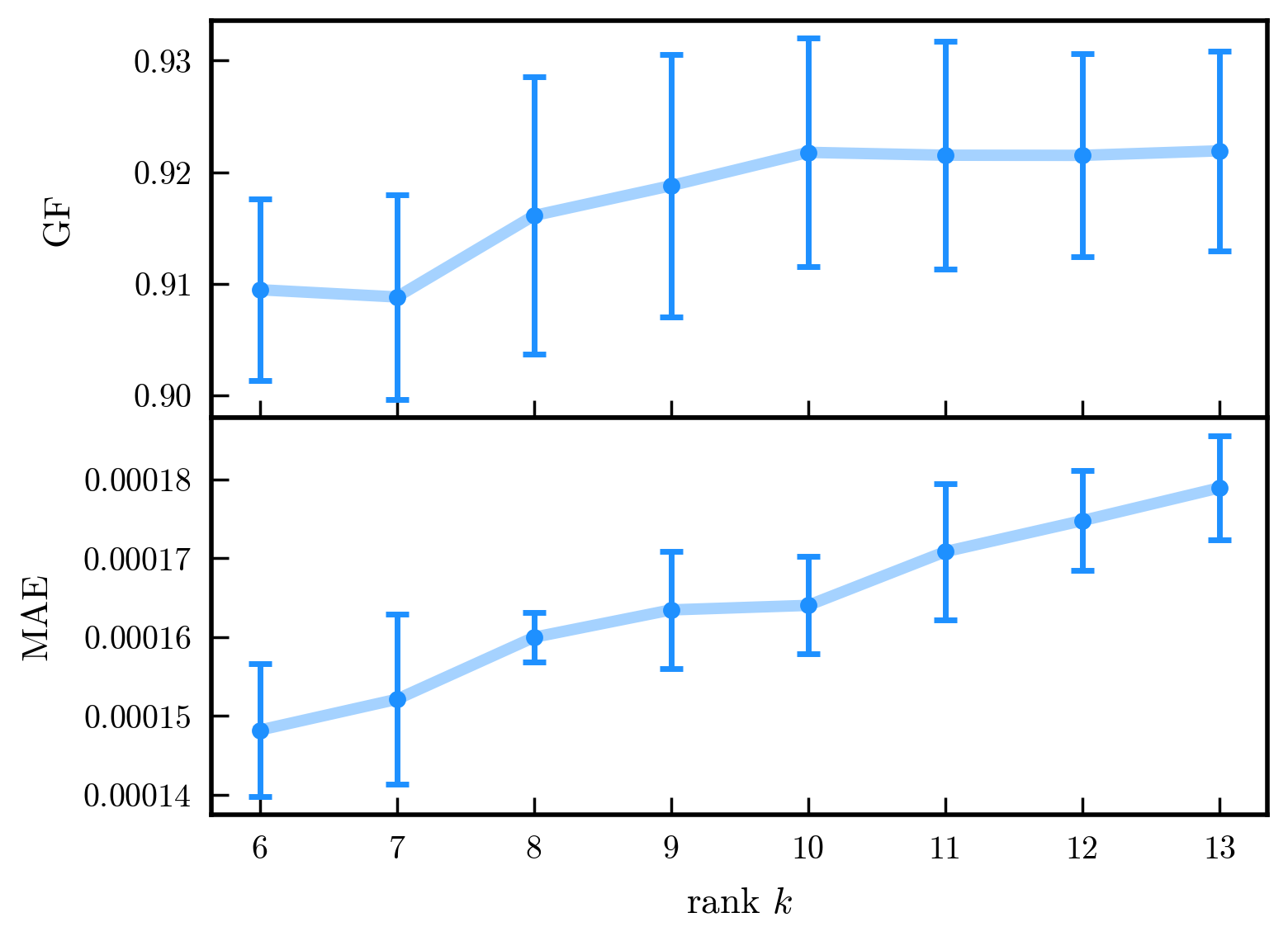}
    \caption{NMF Rank selection. The top and bottom panels show the mean GF and mean MAE with outlier rejection across the test folds, for ranks between 6 and 13. Corresponding standard deviations are shown as error bars. Both metrics are reported for spectra with ZCONF values of 2 and 3.}
    \label{fig:xval}
\end{figure}

Given that our spectra matrix $X$ is large ($9252 \times 53918$), we chose $K=5$; this number is a good trade-off between computational speed, bias, and variance. We used the GF and MAE to quantify redshift prediction performance on the test folds. Results of this exercise are reported in Fig.~\ref{fig:xval}. 

We observe an improvement in the GF up to rank $10$, beyond which it plateaus. In contrast, the MAE (with outlier rejection) increases steadily with rank, indicating that while a larger rank does not degrade the GF, it leads to less precise redshift estimates for non-outliers. Based on these observations, we fix the rank at $10$ for the remainder of this work, as it gives the highest success rate and more precise redshift estimates compared to higher ranks.

\begin{figure*}[b]
    \centering
    \includegraphics[width=1\linewidth]{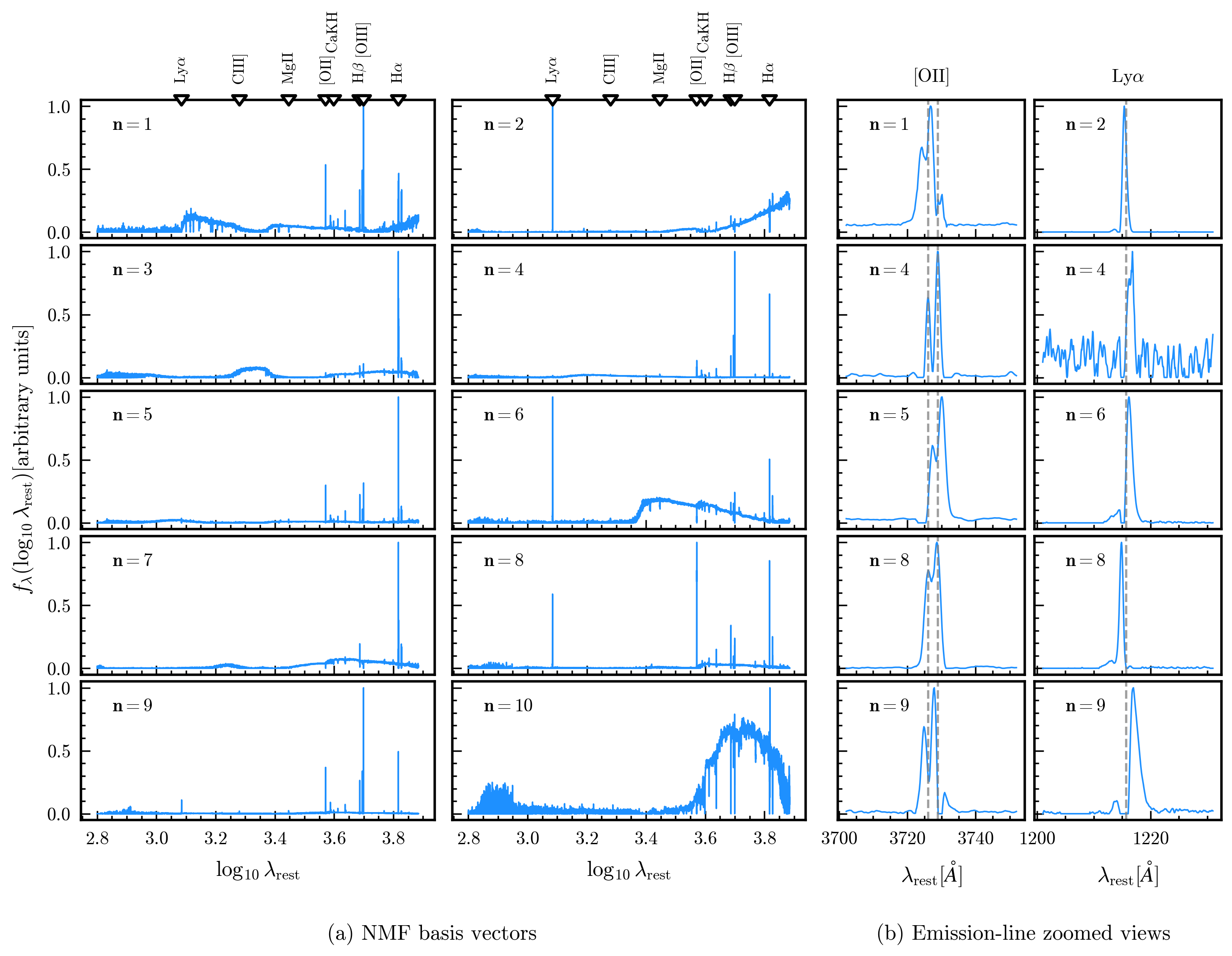}
    \caption{NMF learned basis vectors. Panel (a) shows the basis vectors obtained from a sequential rank–10 NMF decomposition applied to 80\% of the sample. The index $n$ labels each basis vector. The names and rest-frame wavelengths of prominent spectral emission and absorption lines are indicated in the upper panels. Panel (b) presents zoomed-in views of the \mbox{[\ion{O} \textsc{ii}]} and Ly$\alpha$ emission lines, shown in the first and second columns, respectively. In each zoom-in, the line's peak flux is normalized to one, the line rest wavelength is shown with a vertical dashed gray line, and the corresponding basis vector is indicated.}
    \label{fig:nmf_basis_vectors}
\end{figure*}

To conclude this section, we present in Fig.~\ref{fig:nmf_basis_vectors}, the basis vectors obtained from a rank–10 NMF decomposition, applied to 80\% of our sample. The basis vectors capture a broad range of spectral features, including prominent emission and absorption lines such as Ly$\alpha$, \mbox{[\ion{O}\textsc{ii}]}, CaKH, H$\beta$, \mbox{[\ion{O}\textsc{iii}]}, and H$\alpha$, with varying relative line ratios across components. In addition, they encode distinct stellar continuum characteristics: for example, vector~10 is associated with red, evolved galaxy populations. Vector~1 shows a blue continuum with a clear UV spectral slope indicative of star-forming galaxies.
\noindent We further note that several emission lines, in particular \mbox{[\ion{O} \textsc{ii}]} (see panel~b), do not always coincide with their nominal rest-frame wavelengths (indicated by the vertical gray lines). Instead, the line center appears blue-shifted in vectors~1 and~9, redshifted in vector~5, and consistent with the rest frame in the remaining vectors. Similar behavior is observed for Ly$\alpha$, which additionally exhibits a blue bump in some components. These variations indicate that the NMF basis naturally captures kinematic and radiative transfer effects present in the galaxy spectra.

\begin{figure*}[t]
    \centering
    \includegraphics[width=0.9\linewidth]{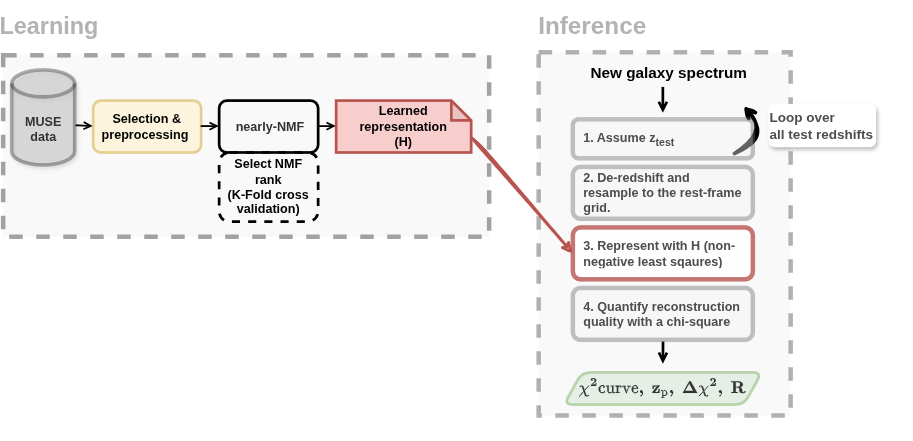}
    \caption{The workflow of redshift prediction using the rest-frame representation learned with nearly-NMF}
    \label{fig:nmf_chart}
\end{figure*}

A schematic overview of the main steps of our method is provided in Fig.~\ref{fig:nmf_chart}.

\section{Results}
\label{sec:application}

\subsection{Performance results}
\label{subsec:performance_results}
We evaluated the performance of our method using the first fold of the K-fold cross-validation setup. Rank-10 NMF basis vectors were derived from the learning subset, and performance was assessed on the testing subset, ensuring that no spectral data from the test set influenced the learning process. This evaluation focused exclusively on \num{1454} spectra with a ZCONF $\geq2$.

\begin{figure}[h]
    \centering
    \includegraphics[width=0.7\linewidth]{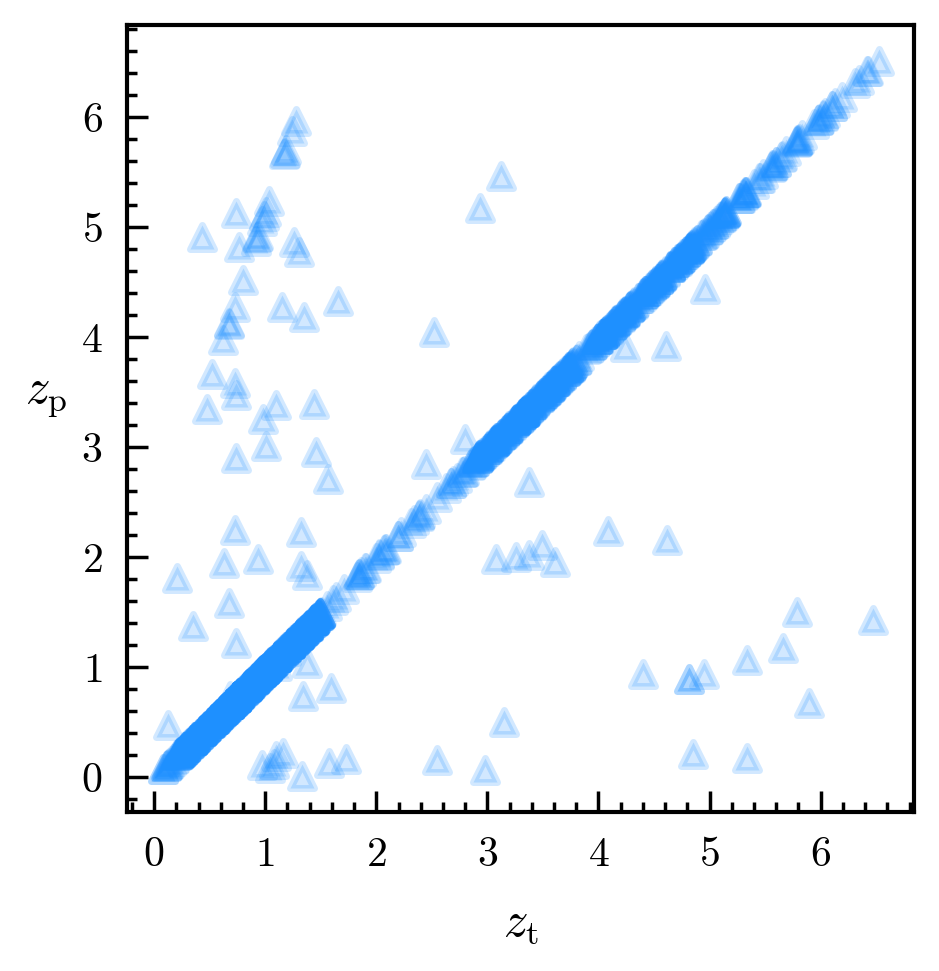}
       \caption{Scatter plot of the predicted redshift ($z_p$) against true redshift ($z_t$) for the test set.}
    \label{fig:scatter}
\end{figure}

Figure~\ref{fig:scatter} shows the scatter of the predicted redshift ($z_p$) against the true redshift ($z_t$). Our method successfully recovers the redshift for \num{1363} spectra (points on the diagonal), while \num{91} spectra have bad predictions (off-diagonal points), yielding an overall GF of \num{93.7}$\%$. Inspection of the failures reveals three main causes: (1) spectra with poor flat-fielding, characterized by distorted continua at the blue and red ends; (2) spectra with a large fraction of negative flux values (negative median flux); and (3) confusion between the \mbox{[\ion{O}\,\textsc{ii}]} doublet and  Ly$\alpha$, especially in spectra lacking stellar continuum and/or with broadened \mbox{[\ion{O}\,\textsc{ii}]} profiles due to galaxy kinematics. The latter confusion appears as a secondary diagonal above the main one in the scatter plot.

\begin{figure}[h]
    \centering
    \includegraphics[width=0.98\linewidth]{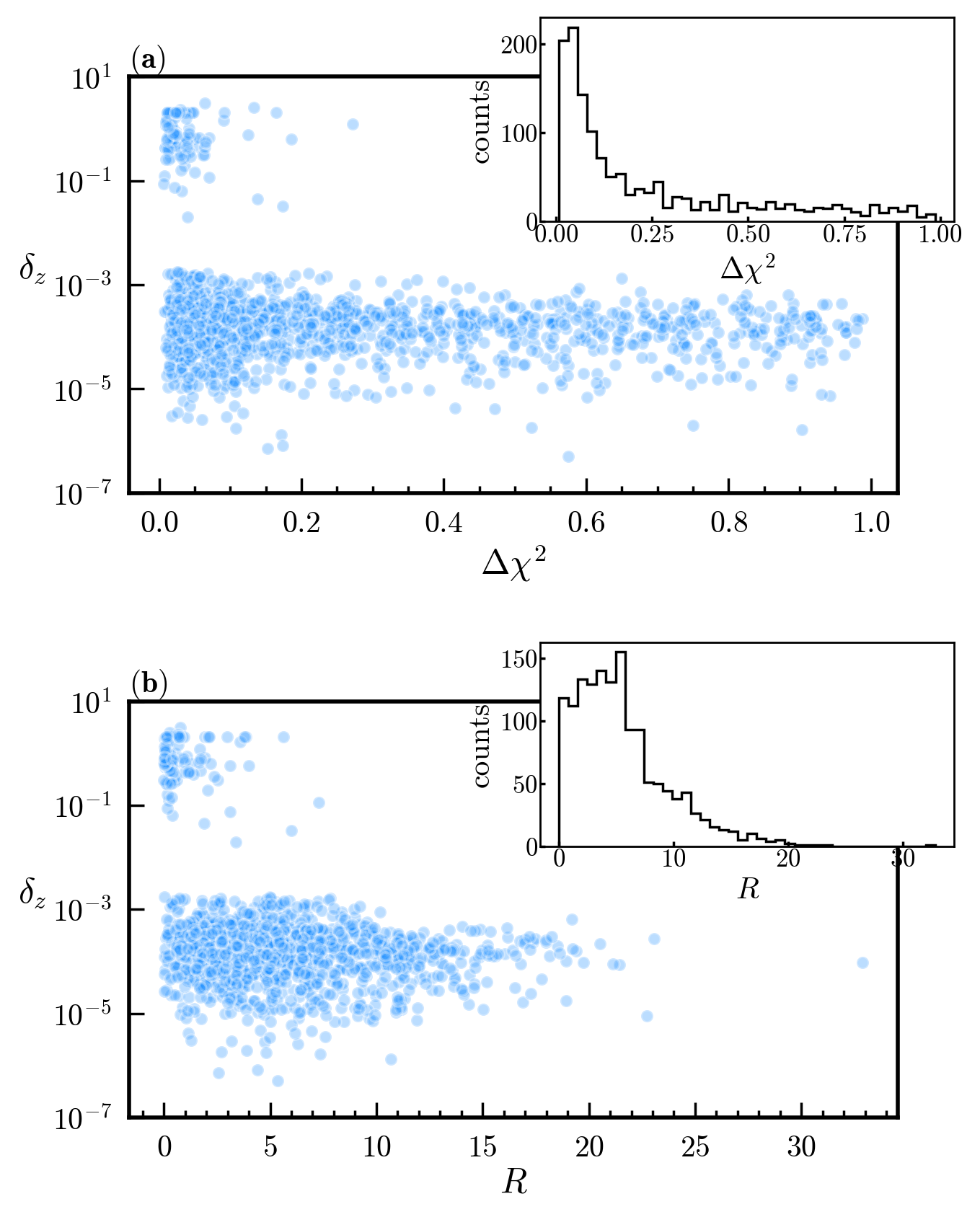}
       \caption{Redshift prediction error and significance. (a) Scatter plot of redshift relative error ($\delta_z$) against significance score $\Delta \chi^2$ for the test set. (b) Scatter plot of $\delta_z$ against Robustness score $R$.}
    \label{fig:scatters}
\end{figure}

Figure~\ref{fig:scatters} presents the relative error $\rm \delta_{z}$ as a function of the significance score $\Delta \chi^2$ (panel (a)) and the Robustness metric $R$ (panel (b)). We observe that almost all failed predictions correspond to $\Delta \chi^2 < 0.05$ and $R < 3$. Moreover, we note that the $R$ distribution peaks at 5 and failed predictions cluster on the left tail of this distribution; a cut in $R$ can therefore be used to secure redshifts (pure selection). Cases with $R>3$ but incorrect predictions are associated with spectra dominated by negative flux values.  

\begin{figure}
    \centering
    \includegraphics[width=0.98\linewidth]{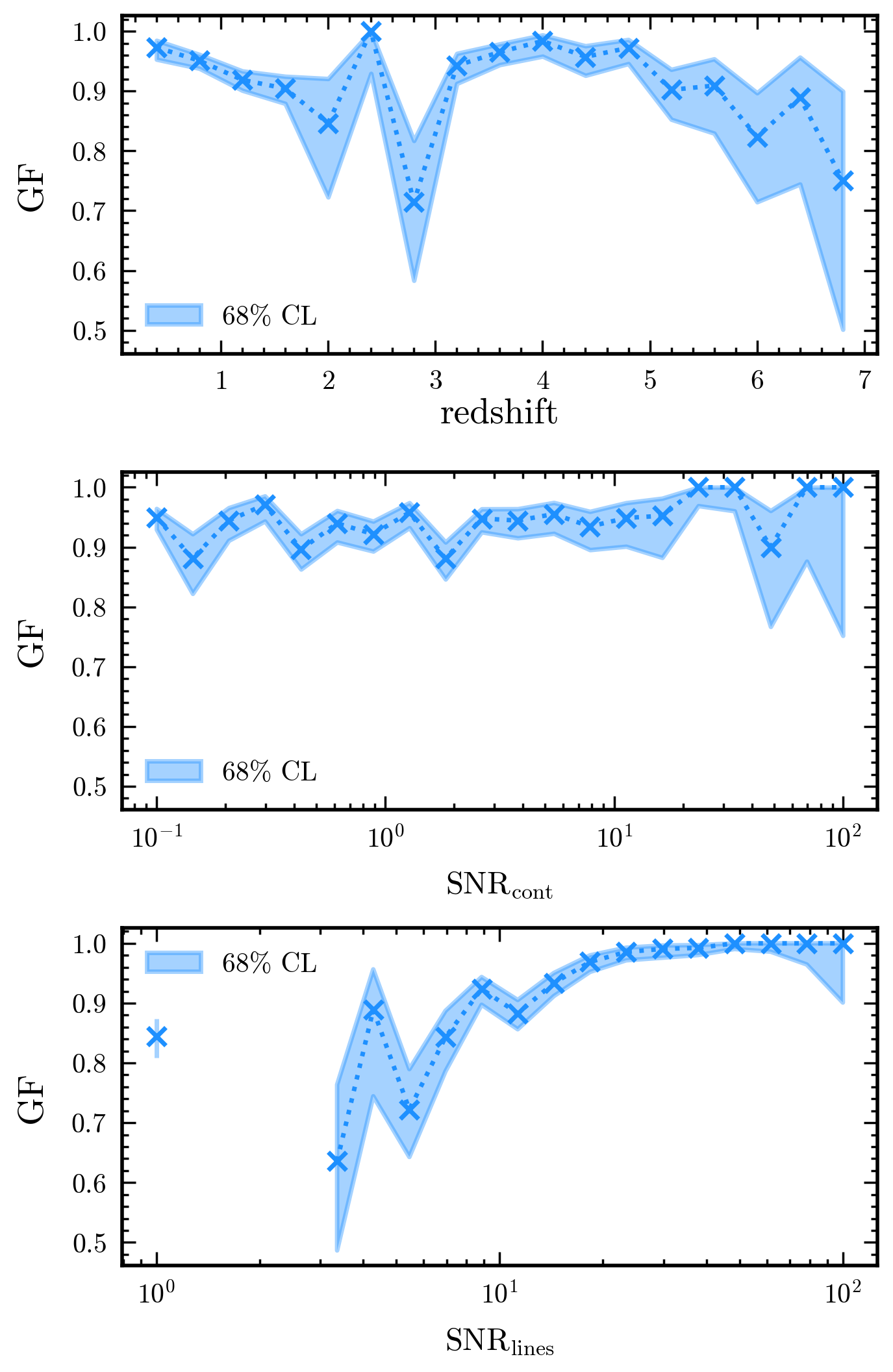}
    \caption{Performance as a function of redshift and SNR. (a) Average GF as a function of redshift, (b) GF as a function of the continuum SNR ($\rm SNR_{cont}$), (c) GF as a function of the lines SNR ($\rm SNR_{lines}$). The shaded regions indicate 68\% Wilson confidence intervals.}
    \label{fig:test_snr}
\end{figure}

The three panels in Fig.~\ref{fig:test_snr} present the average GF as a function of redshift, continuum SNR, and lines' SNR. Overall, the GF remains above 90\% across most redshift bins, with notable declines in the redshift desert ($z \sim 2$ and $z \sim 2.8$) and at very high redshift ($z \geq 6$). In the redshift desert, the reduced performance is due to the lack of strong spectral features in the spectra. At the highest redshifts, the GF has large uncertainties due to the small number of objects in these bins.
When examined as a function of continuum SNR, the GF is largely stable and remains above 90\% for nearly all bins, exhibiting only mild fluctuations at low SNR. A modest performance improvement is observed toward higher continuum SNR values, although the dependence is comparatively weak.
In contrast, the strongest dependence is seen as a function of the lines' SNR. The GF is very low for $\mathrm{SNR}_{\rm lines} \lesssim 7$, rises steadily with increasing SNR, and approaches unity at $\mathrm{SNR}{\rm lines} \sim 13$, beyond which it plateaus. This trend indicates that the number of emission/absorption lines and their signal-to-noise ratios are the dominant factors governing the GF.

\subsection{Performance with varying depth}
\begin{figure}
    \centering
    \includegraphics[width=1\linewidth]{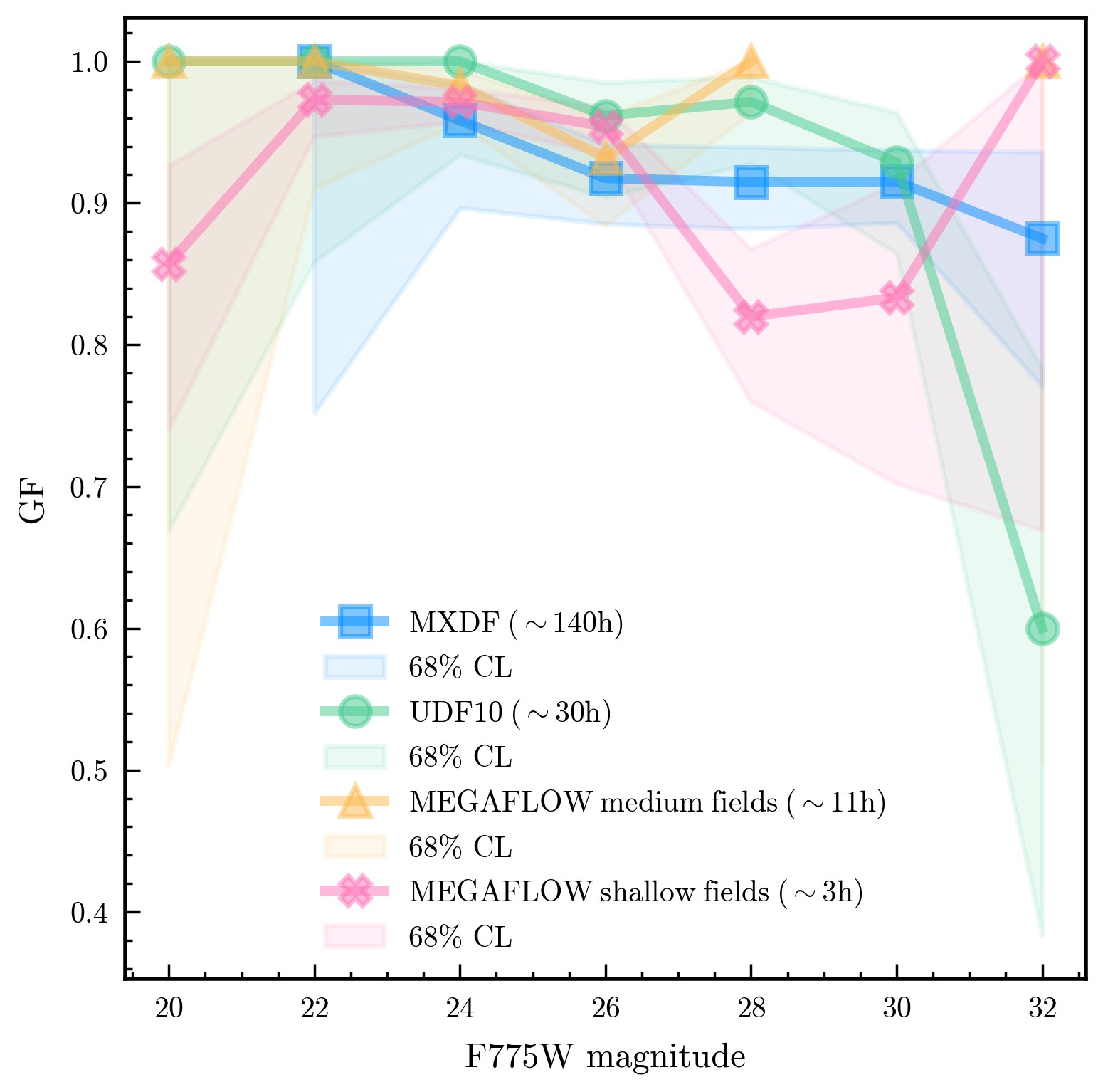}
    \caption{Performance with varying depth. The different curves show the averaged GF in bins of the $F775W$ magnitude for various surveys with different depths. Shaded regions indicate 68\% Wilson confidence intervals.}
    \label{fig:ginofli}
\end{figure}

Increasing survey depth improves data quality by reducing noise and enables the detection of faint objects. To investigate the impact of exposure time on our method, we applied it to four samples of spectra with different exposure times: MXDF ($\sim$\,140 hours), UDF10 ($\sim$\,30 hours), MEGAFLOW medium fields ($\sim$\,11 hours), and MEGAFLOW shallow fields ($\sim$\,3 hours). We then measured the average GF in bins of the $F775W$ magnitude for each sample; as shown in Fig.~\ref{fig:ginofli}.\\ 

As expected, the GF improves with increasing survey depth. The MXDF sample delivers the best performance, maintaining a GF $\gtrsim 90\%$ down to $F775W \sim 32$. The UDF10 and MEGAFLOW medium fields show comparable performance and perform only worse at magnitudes fainter than $\sim 30$; for the MEGAFLOW medium fields, this regime is affected by low-number statistics. In contrast, the MEGAFLOW shallow fields already show a decline in GF below 90\% at $F775W \sim 28$. These results demonstrate that survey depth is a key factor for achieving reliable redshift predictions at faint magnitudes ($F775W \gtrsim 28$).

\subsection{Threshold on \texorpdfstring{$\mathrm{\Delta \chi^2}$}{Lg}}
\label{subsec:threshold}
Unlike MOS, where spectroscopy is carried out on preselected targets (e.g., photometrically selected), IFS provides spectroscopy of everything in the field of view. 
In MUSE surveys,  sources must be blindly detected using algorithms such as ORIGIN \citep{Mary_2020} or FELINE \citep{Wendt2025}, which can produce many false positives in order to be highly complete. Currently, these false sources are processed by the redshift prediction pipeline just like real objects, and identified and marked as ZCONF0 during the visual inspection step \citep[e.g.][]{BaconR2023}.

A reliable automatic redshift prediction tool must cope with these contaminants and classify them accordingly. Here, we investigate the ability of our method to discriminate between true and false sources, relying on the $\rm \Delta \chi^2$ significance score. Specifically,
we applied our redshift prediction method to ZCONF0 spectra from the MXDF survey, and then compared their $\rm \Delta \chi^2$ predicted redshift significance score distribution with those of spectra having higher ZCONF scores taken from the same survey. 

\begin{figure}
    \centering
    \includegraphics[width=1\linewidth]{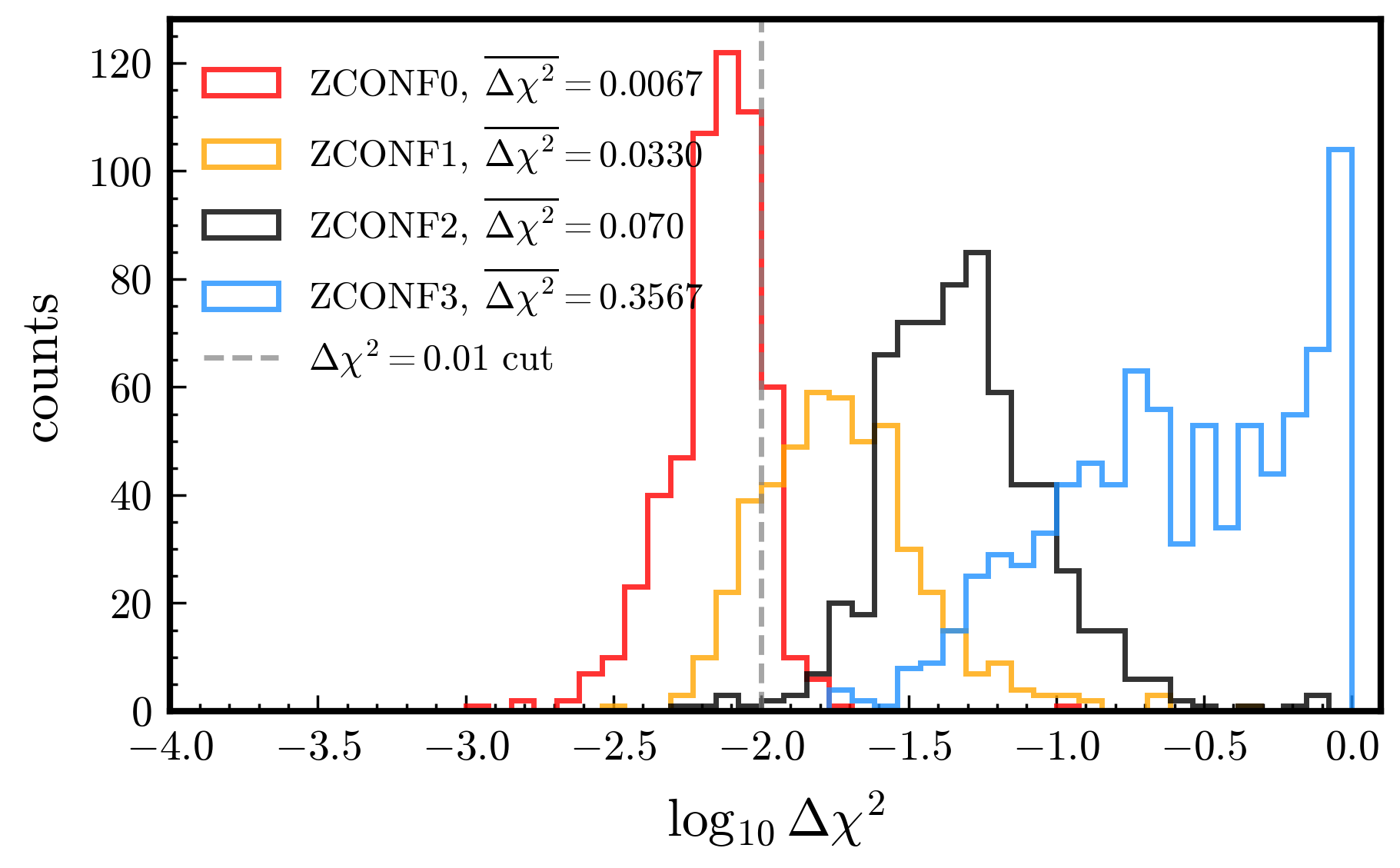}
    \caption{Redshift significance score and ZCONF. $\log_{10} \Delta \chi^2$ distribution for the 4 classes of ZCONF in the MXDF survey.}
    \label{fig:zscore_dist}
\end{figure}

Figure~\ref{fig:zscore_dist} shows the distribution of $\rm \log_{10} \Delta \chi^2$ for the $4$ ZCONF classes. Looking at the four distributions, we can see that the significance score discriminates ZCONF2 and ZCONF3 classes from the ZCONF0 class, with almost no overlap. We note that ZCONF2 sources that overlap with ZCONF0 originate from spectra with a high fraction of negative values. ZCONF1 distribution, on the other hand, shows some overlap with ZCONF0 distribution. In this overlapping region, it becomes hard to distinguish between a very faint true source and a false one, e.g., originating from bad flat-fielding.

We observe that a threshold at $\log_{10} \Delta \chi^2 =-2$, allows us to keep all of the ZCONF2 and ZCONF3 sources and a large fraction of ZCONF1 sources while rejecting most of the ZCONF0 sources. For this cut, we quote a completeness score\footnote{Fraction of ZCONF $1$, $2$ or $3$ sources that are successfully selected} of $95.9\%$, and a purity score\footnote{Fraction of selected sources that are truly ZCONF $1$, $2$ or $3$} of $96.0\%$.

\subsection{Application on blends}
To ensure the robustness of our analysis, we have so far deliberately excluded blended sources. For large-scale applications and studies involving deep fields, however, it becomes essential to identify and flag such blended sources, as they are ubiquitous \citep{2021NatRP...3..712M} and constitute a primary cause of incorrect redshift determinations, given that redshift pipelines (including ours) generally recover the redshift of the most luminous component \citep{10.1093/mnras/stw3021}.

\begin{figure*}
    \centering
    \includegraphics[width=0.9\linewidth]{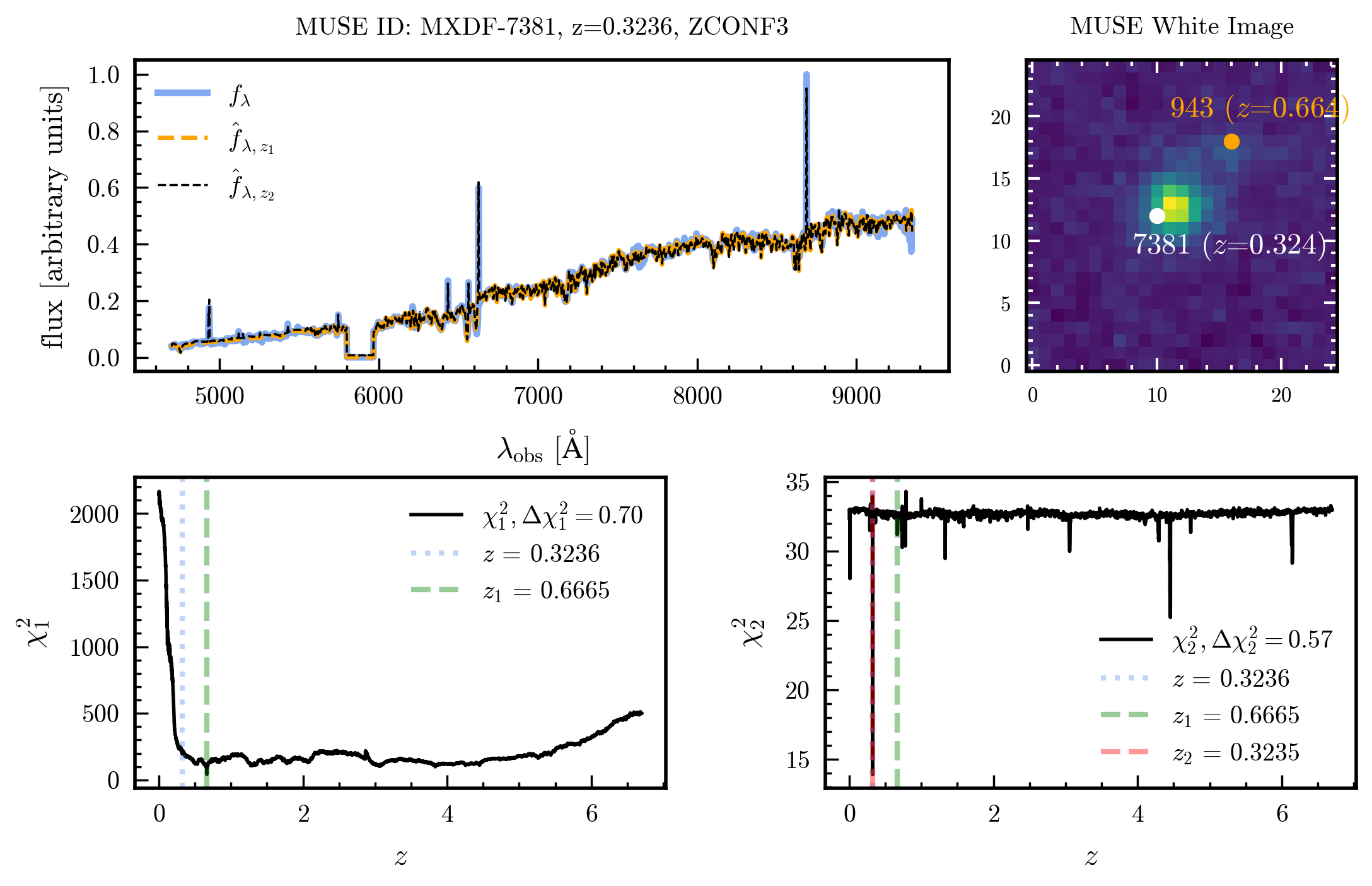}
    \caption{Debelending MXDF-7381. The top left panel shows the spectrum of MXDF-7381 ($f_\lambda$) in blue, best fit at $z_1$ ($\hat{f}_{\lambda, z_1}$) in orange and the second reconstruction at $z_2$ ($\hat{f}_{\lambda,z_2}$)  as a dashed black line. Bottom panels show the obtained $\chi^2_1$ and $\chi^2_2$ curves from left to right. $\chi^2_1$ reveals a minimum at a redshift $z_1 = 0.6665$ corresponding to the blending component, whereas $\chi^2_2$ reveals a minimum at $z_2 = 0.3235$ which corresponds to MXDF-7381 true redshift. The top right panel shows a white image of MXDF-7381 with nearby sources, where we can see the presence of a close source, MXDF-943, which has the blending redshift $z_1$.}
    \label{fig:debelending_eg}
\end{figure*}

In this subsection, we explore the feasibility of blend detection based exclusively on 1D spectra using our learned NMF basis vectors. We follow the framework described in \citet{TsalmantzaHogg2012a} with some adaptations. The method proceeds as follows:
\begin{enumerate}[(a)]
    \item For a given spectrum $\vec{f_{\lambda}}$, we predict its redshift using our method. We note the obtained $\chi^2(z)$ curve $\chi^2_1$ and the corresponding redshift $z_1$. Note that if the spectrum is blended, this step identifies the redshift component exhibiting the strongest features.
    \item We use $\texttt{pyPlatefit}$ to fit the spectrum at $z_1$, and note the fitted spectrum $\vec{\hat{f}_{\lambda,\,z_1}}$. The goal of this step is, on the one hand, to maximally reduce reconstruction residuals, and on the other hand, to fit rare lines that are not captured by our NMF basis vectors. If not accounted for, these rare lines result in false blend detections. To this end, we extended the default line list used by $\texttt{pyPlatefit}$ to include $\ion{Fe}\,\textsc{ii}^\star$, $\ion{C}\,\textsc{ii}^\star$ and $\ion{Si}\,\textsc{ii}^\star$ fluorescence emission lines, $\ion{He}\,\textsc{i}\,\lambda3187.745$, $\ion{He}\,\textsc{i}\,\lambda4471.479$, $\ion{He}\,\textsc{i}\,\lambda5875.624$, and $\mbox{[\ion{N}\,\textsc{i}]}\,\lambda5200.257$ emission lines.
    \item Following the same steps in our redshift prediction method, we scan all possible redshifts a second time. This time at each test redshift, we augment our basis by injecting $\vec{\hat{f}_{\lambda,\,z_1}}$. We denote by $\bm{H}^{(z_i)}$ the NMF basis vectors corresponding to redshift $z_i$,  where $i$ indexes test redshifts. The augmented basis at $z_i$ is written as
    \begin{equation*}
        \bm{H}'^{\; (z_i)} = \left\{ \vec{\hat{f}_{\lambda, z_1}} , \bm{H}^{(z_i)} \right\}.
    \end{equation*}
    At each $z_i$, we non-negatively project $\vec{f_\lambda}$ into $\bm{H}'^{\; (z_i)}$, and compute the reconstruction error, following the same equations in \ref{eq:nmf_reconstruction} and \ref{eq:nmf_chi2test}. We note the obtained $\chi^2(z)$ curve $\chi^2_2$.
    \item We identify all minima in $\chi^2_2$ and discard minima that are in confusion with $z_1$. If a minimum with $\Delta \chi^2 > t_{\rm blend}$ remains---$t_{\rm blend}$ is a threshold for class separation---, we flag the spectrum as blended. The remaining minimum with the highest significance score  $\Delta \chi^2$ gives the blending redshift $z_2$; we note the second reconstruction of the NMF $\vec{\hat{f}_{\lambda,\,z_2}}$.
\end{enumerate}
These steps can be repeated to search for double or even triple blends. Here, we stop after one blend detection trial, as our goal is to detect blends and not to estimate the number of mixing components. Figure~\ref{fig:debelending_eg} illustrates an example of a blended source detected by our approach.

We applied our deblending approach to all MXDF survey sources with a ZCONF of 1 or higher, for a total of 664 objects. We inspected all sources and flagged blended ones using the $\texttt{SourceInspector}$ software and the \texttt{AMUSED}~\footnote{\url{https://amused.univ-lyon1.fr/}} interface \citep{BaconR2023}~\footnote{\texttt{SourceInspector} enables interactive inspection of narrow-band images and localization of emission within the datacube, while \texttt{AMUSED} provides an efficient way to identify nearby sources}. From the 664 objects, we flagged 183 as blended and removed 3 objects classified as stars.

\begin{figure}[h]
    \centering
    \includegraphics[width=0.95\linewidth]{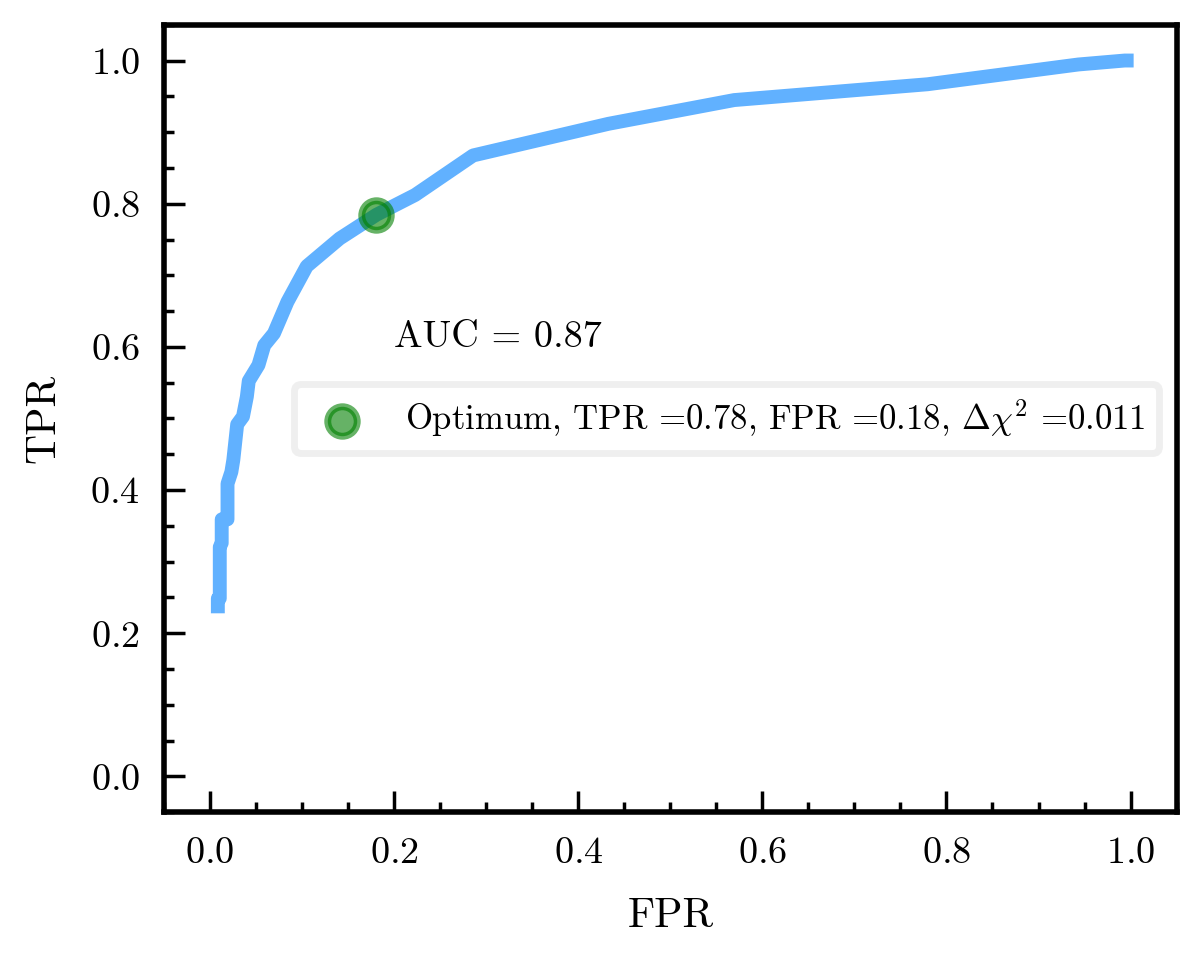}
    \caption{Deblending test ROC curve. The TPR versus the FPR is plotted as a blue line. The value of the AUC score is shown, and the optimal threshold is marked by the green point.}
    \label{fig:roc}
\end{figure}

For this test, we present the Receiver Operating Characteristic curve (ROC curve) in Fig.~\ref{fig:roc}. This curve reports the true positive rate (TPR) versus the false positive rate (FPR) for a series of thresholds ($t_{\rm blended}$ ) spanning 0.003-0.05 with a step of 0.001. TPR and FPR are defined as
\begin{equation*}
      \rm TPR = \dfrac{TP}{P}, \;\;FPR =\dfrac{FP}{N},
\end{equation*}
where TP is the number of true positive detections (correctly predicted as blended), FP is the number of false positive detections (incorrectly predicted as blended), P is the total number of positive labels (number of blends), and N is the total number of negative labels (number of sources not blended). Next, we quantify the performance based on the ROC curve and choose a threshold that best balances between the TPR and FPR.

To quantify the performance of our deblending approach, we use the Area Under the Curve (AUC) score. Values close to 1 indicate excellent discrimination, and in our case, the AUC is 0.87, categorizing our method as having "good" performance. 

We selected the optimal threshold as the point that minimizes the Euclidean distance to the ideal classifier (0,1), resulting in a TPR of 0.78, a FPR of 0.18, and a threshold on $\Delta\chi^2$ equal to 0.011. We note that this value is very close to the separation threshold between true and false sources (see \ref{subsec:threshold}).  

Upon examining the classifications, we found that false positives mainly result from sky-subtraction residuals, bad $\texttt{pyPlatefit}$ fits, and noise. Undetected blends at the optimal threshold correspond to very weak blends. Additionally, we observed that some blends remain undetected even at low thresholds ($\Delta\chi^2 < 0.01$). This occurs when a blending with one line coincides with a line of the first redshift, in which case $\texttt{pyPlatefit}$ absorbs the blending line in step (b).

We conclude this subsection by suggesting possible improvements. Unlike \citet{TsalmantzaHogg2012a}, who doubles the basis vectors at each test redshift---keeping one set fixed at the first redshift while the other slides with the test redshift---our approach augments the basis vectors using the \texttt{pyPlatefit} best fit at the first redshift. This strategy has the advantage of minimizing residuals at the location of spectral lines; however, it also results in the continuum being completely absorbed during the fit. Introducing a continuum model with fewer degrees of freedom could mitigate this issue and enable the detection of blended objects via continuum features. Additionally, exploiting the spatial origin of the emission would help identify cases in which a blending with one line coincides with another line of the first redshift.  

\subsection{Application to a new dataset}
We tested our method on a new reduction of the MUSE-WIDE survey (DR2), which benefits from improved sky-subtraction and flat-fielding (Urrutia, in prep), and includes more sources (2332 sources in DR2 compared to 1602 sources in DR1).

After inspecting all 2332 spectra and removing blends and spectra with very broad features, we obtain an overall GF of 97.1\% on ZCONF2 and ZCONF3 classes. The GF of each ZCONF class is reported in Table \ref{tab:musewidedr2}. Inspection of the failed predictions reveals that the majority of them are due to confusion between \mbox{[\ion{O}\,\textsc{ii}]} and Ly$\alpha$.   

\begin{table}[h]
    \centering
    \caption{MUSE-WIDE DR2 survey test results.}
    \begin{tabular}{c|c|c|c}
        ZCONF   & $1$ & $2$ & $3$ \\
        \hline
        GF      & $77.2 \%$ & $95.2\%$ & $99.0 \%$
    \end{tabular}
    \label{tab:musewidedr2}
\end{table}

\subsection{Implementation and computational time}

We make available the data matrix used in this work, implementation of the nearly-NMF algorithm, and the code for our redshift prediction method in a \texttt{Julia} package \href{https://github.com/masten-bourahma/Moose.jl}{\texttt{Moose.jl}}, along with a \texttt{Python} wrapper. Our implementation takes approximately $200\,\rm ms$ to test 7000 redshifts for a single input spectrum; the code was run on a 24-threads CPU with multi-threading applied to parallelize multiple redshift tests. This time can be reduced in many ways: by decreasing the resolution $\delta \Gamma$ of the basis vectors, by selecting redshifts to test if prior information is available, or by the use of a CPU with more threads. Such computation time is reasonable for current MUSE surveys.
\section{Discussion}
\label{sec:discussion}
In this study, we demonstrated that learning a rest-frame representation of galaxy spectra through NMF in a data-driven setup provides an effective framework for automated redshift estimation. By projecting new spectra onto this representation under non-negativity constraints and identifying the best-fitting decomposition, our method attained an overall success rate of $93.7\%$ when applied to MUSE spectra. To further quantify the reliability of each measurement, we introduced a $\Delta \chi^2$ score; adopting a threshold at  $\Delta \chi^2 = 0.01$ yielded a very good separation between ZCONF0 and higher-confidence redshifts, illustrating the robustness of our approach against false detections (ZCONF0), which are common in IFS data. Finally, we assessed the feasibility of blend detection using our NMF basis and showed that approximately 78\% of blended objects can be recovered at a false positive rate of 18\%.

Because standard redshift inference tools have been developed and tested primarily on low-redshift spectra ($z$<3), we do not attempt a direct numerical comparison. Instead, we emphasize that our method was applied to MUSE spectra spanning redshifts between 0 and 6.7, a range that presents significant challenges: line confusion such as Ly$\alpha$/\mbox{[O\,\textsc{ii}]} and Ly$\alpha$/H$\alpha$, and the redshift desert. Achieving an overall $93.7\%$ success rate across this regime demonstrates that our method is robust under conditions that are not systematically addressed by existing tools.

Next, we compare our method with the methodological choices in established approaches. Classical tools, such as \texttt{RedRock}, estimate redshifts by fitting a set of archetype templates constructed from a combination of observed and synthetic spectra, whereas our method learns the templates directly from the data, enabling it to adapt to the diversity of the sample.
\texttt{AUTOZ}, on the other hand, relies on cross-correlation with a set of SDSS templates, which requires subtracting the stellar continuum and discards physically informative broadband features. Cross-correlation also suffers from well-known issues, including template mismatch, ambiguous peaks in the cross-correlation function when a spectrum is dominated by a single emission line, and reduced robustness when spectra deviate from the assumed template family. In comparison, our method leverages the full stellar continuum, which constitutes a discriminative source of information, especially when separating between Ly$\alpha$ and \mbox{[\ion{O}\,\textsc{ii}]}. Recent deep-learning approaches, such as $\texttt{GaSNet III}$ and $\texttt{M-TOPnet}$, train neural networks to predict redshifts using strategies such as multi-task learning and representation learning. While these networks achieve high performance, they require large labeled datasets (> \num{20000} spectra) for training, and typically operate on continuum-subtracted inputs, which again removes potentially discriminative information. By contrast, our method requires far fewer training examples ($\lesssim$7,000) and processes the full spectrum, making it both competitive with and complementary to these approaches.

Despite its robustness, our method has limitations. We find reduced performance in spectra affected by artifacts such as: (1) residuals from imperfect sky subtraction, (2) flat-fielding errors that distort the continuum, and (3) spectra with a negative median flux, where NNLS decompositions often fail. Because such artifacts are unavoidable in IFS data, a promising direction is to replace the NNLS solver with a neural network (e.g., encoder-like architecture) that enforces non-negativity while learning to ignore such artifacts, potentially leading to further performance gains.
\section{Conclusions}
\label{sec:conclusions}
In this work, we presented a method for automatic redshift prediction for MUSE galaxy spectra. These spectra span a wide redshift range ($0 < z < 6.7$), where classical tools perform poorly or have not been systematically tested. The method learns a low-rank representation through non-negative matrix factorization and estimates redshifts by projecting new spectra onto this basis and evaluating $\chi^2$ errors across trial redshifts.

Using a K-fold cross-validation, we identified ten basis vectors as the optimal representation, and our method achieved a Good Fraction of $93.7 \%$ on an independent test set. Most of the failed predictions resulted from confusion between the \mbox{[\ion{O}\,\textsc{ii}]} doublet and Ly$\alpha$. To assess the reliability of a prediction, we introduced two metrics: the significance score $\Delta \chi^2$ and the robustness score $R$. We found that failed predictions are typically associated with low values of both metrics, with $R$ showing good discriminating power.

We demonstrated an application of our method to spectral deblending. The approach first identifies and fits the dominant redshift component, then re-scans for secondary redshift solutions after augmenting the NMF basis with the fitted spectrum. When applied to MXDF sources, the method achieved strong performance, recovering 78\% of blended sources with a false positive rate of 18\%.

Applied to a dataset with improved reduction, the method reached a Good Fraction of $97.1 \%$ on sources with ZCONF2 and ZCONF3, demonstrating robust generalization and sensitivity to data quality. We further evaluated its ability to identify false detections and showed that a threshold at $\Delta \chi^2 = 0.01$ efficiently separates genuine from spurious sources.

Overall, these results demonstrate that low-rank representations obtained via NMF provide a flexible and data-driven framework for automated redshift determination across a broad redshift range. In addition to high accuracy, the method offers intrinsic reliability diagnostics, making it well-suited for application to current and coming large spectroscopic surveys.
\begin{acknowledgements}
     We thank A. Casey for fruitful discussions in the initial phases of the project.
     NB acknowledges support from the ANR DARK grant (ANR-22-CE31-0006).
     MB acknowledges support from the Blaise Pascal GPU computing infrastructure.     TU acknowledges funding from the European Research Council (ERC) under the European Union's Horizon 2020 research and innovation programme (grant agreement 101020943, SPECMAP-CGM).
     Part of this work was supported by the German \emph{Deut\-sche For\-schungs\-ge\-mein\-schaft, DFG\/} project number Ts~17/2--1.\\
\end{acknowledgements}
\bibliographystyle{bibtex/aa.bst}
\bibliography{references.bib} 
\end{document}